\documentclass[aps,prb,reprint,superscriptaddress,amsmath,amssymb,floatfix]{revtex4-2}
\usepackage{graphicx}
\usepackage{dcolumn}
\usepackage{bm}
\usepackage{hyperref}
\usepackage{xcolor}
\usepackage{pgfplots}
\usepackage{pgfplotstable}
\usepackage{filecontents}
\usepackage{listings}
\usepackage{booktabs}
\usepackage{tikz}
\usepackage{subcaption}
\pgfplotsset{compat=1.18}
\graphicspath{{figures/}}

\hypersetup{colorlinks=true,citecolor=blue,linkcolor=blue,urlcolor=blue}
\newcommand{\dd}{\mathrm{d}}
\newcommand{\ee}{\mathrm{e}}

\newcommand{\Ltwo}{L^2}
\newcommand{\Linf}{L^\infty}

\begin{document}

\title{A Reproducible Demonstration of AI-Assisted Scientific Workflow on Canonical Benchmarks}

\author{Kin Hung Fung}
\email{funguiuc@gmail.com}

\begin{abstract}
We present a fully reproducible demonstration of an AI-assisted scientific workflow designed for a broad physics, mathematics, and computer-science readership. The initial project artifact stack was generated from one single user prompt and then reviewed and curated for submission by the human author. Rather than claiming a new scientific discovery, the manuscript uses canonical benchmark problems with exact, manufactured, or independently checkable answers. The analytical component starts from the one-dimensional quantum harmonic oscillator, derives its dimensionless form, and validates finite-difference eigenpairs against exact Hermite-function benchmarks. The numerical partial-differential-equation component solves a heat equation with a known modal solution and a Poisson problem verified by a manufactured solution, with explicit convergence studies. The inverse-modeling component fits synthetic damped-oscillation data by nonlinear least squares and quantifies parametric uncertainty by bootstrap resampling. The computational-science component compares dense and sparse eigensolvers and contrasts direct and iterative sparse linear solvers, with careful interpretation of machine-dependent timing data. Taken together, the results show that contemporary AI can already serve as a useful scientific copilot for derivation, implementation, validation, visualization, and manuscript preparation, provided that each stage is constrained by benchmark theory, explicit verification, and transparent artifacts. The demonstration is therefore relevant not because the underlying science is novel, but because it offers a concrete template for trustworthy AI use in technical research practice.
\end{abstract}

\maketitle

\noindent\textbf{Authorship and AI-use disclosure.} This submission is authored by a human author, who takes responsibility for its contents. The initial manuscript, code, data, figures, and artifact stack were generated with substantial assistance from a generative AI system from one single user prompt specifying the scientific scope, validation requirements, deliverables, and writing constraints. The present submission version includes subsequent human-reviewed front-matter and disclosure revisions made for publication readiness.

\section{Introduction}
Modern scientific work is not a sequence of isolated tasks. Derivation, simulation, fitting, verification, data visualization, and writing usually interact throughout a single project, and errors in any one stage can silently contaminate the others. For that reason, a useful scientific assistant is not merely a text generator; it must help organize a workflow in which equations, code, data, figures, and prose remain mutually constrained.

Reproducible computational science has long emphasized that trustworthy numerical work requires explicit validation, transparent code, and clear provenance of every reported result \cite{Peng2011,Wilson2014,Sandve2013}. In parallel, the literature on AI for science has documented rapidly growing use of machine learning and large language models across scientific tasks, including literature synthesis, model construction, coding, and experiment design \cite{Wang2023,Zhang2025}. At the same time, the reliability limits of present-day language models remain substantial, particularly when reasoning or self-correction is not externally checked \cite{Huang2024}.

A methodological gap follows from these two facts. Discussion of AI in science is often either anecdotal or promotional, whereas discussion of reproducibility is often detached from present-day AI tooling. What is still comparatively rare is a technically serious, end-to-end demonstration in which AI assistance is judged against canonical benchmarks with known answers, exact solutions, manufactured solutions, convergence studies, and explicit artifact generation.

This paper fills that gap by assembling a complete mini-workflow around standard problems: the one-dimensional harmonic oscillator, the one-dimensional heat equation, a two-dimensional Poisson problem, a nonlinear damped-oscillator inference task, and simple algorithmic scaling comparisons. Each problem is conventional. The point is not to report new science, but to demonstrate that AI can already accelerate and structure substantial parts of scientific work when the task is framed in a validation-heavy manner.

This framing matters. A discovery paper can sometimes tolerate exploratory ambiguity in early stages. A demonstration paper of the present kind cannot. Its standards must instead be conservative: known benchmarks, quantitative error reporting, independent checks, fixed random seeds, explicit scripts, generated figures, and a reproducibility manifest. The results below should therefore be interpreted as a demonstration of workflow capability rather than a claim of autonomous scientific discovery.

\section{Scope, philosophy, and reproducibility standards}
In this paper, ``useful AI'' means something operationally narrow. It means assistance with symbolic manipulation, operator assembly, solver implementation, plotting, and manuscript drafting in a workflow whose intermediate outputs are independently validated. It does \emph{not} mean that the AI system is credited with scientific judgment, epistemic warrant, or autonomous discovery.

Accordingly, every case study in the paper is benchmarked against one of four standards: an exact analytic solution, a manufactured solution, a convergence study, or an independent numerical cross-check. The manuscript, code, data files, figures, and verification report are generated as a single artifact stack by a top-level driver script. In the present demonstration, that stack was initiated from one single user prompt that specified the full technical and stylistic requirements. This design follows the spirit of reproducible computational-science practice \cite{Wilson2014,Sandve2013,Peng2011} while making the role of AI assistance explicit.

\begin{figure*}[t]
    \centering
    \includegraphics[width=0.98\textwidth]{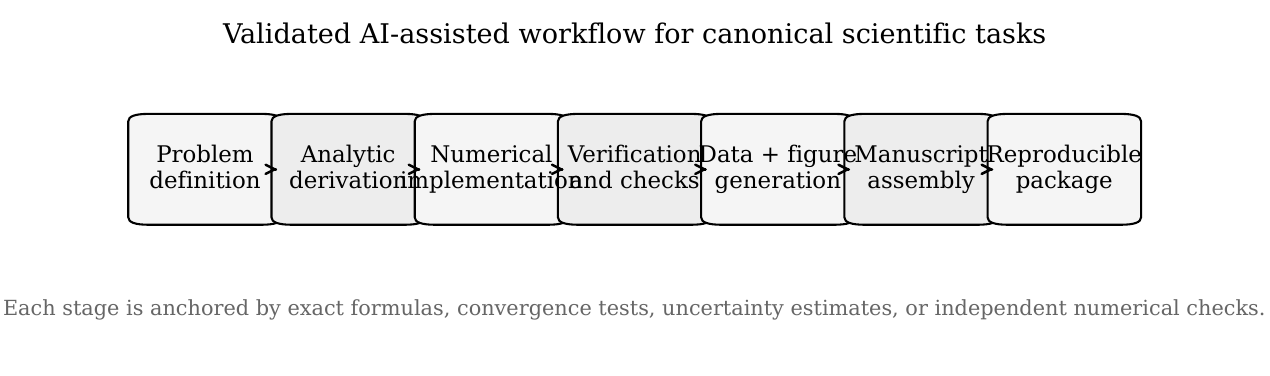}
    \caption{End-to-end structure of the demonstration workflow. The boxes denote the main stages of the project: problem definition, analytic derivation, numerical implementation, verification, data and figure generation, manuscript assembly, and reproducible packaging. The arrows indicate that the workflow is not merely sequential; benchmark formulas and numerical checks constrain downstream coding, and failed checks feed back to earlier stages. The figure is deliberately sober rather than infographic-like because the paper's claim concerns workflow discipline, not interface design.}
    \label{fig:workflow}
\end{figure*}

Figure~\ref{fig:workflow} summarizes the workflow. The central claim of the paper is not that AI removes the need for verification. It is the opposite: contemporary AI is already highly useful precisely when it is embedded in a research protocol that treats verification as a first-class object.

\section{Case study I: symbolic analysis and spectral validation}
\subsection{Benchmark problem and exact reference}
We first consider the dimensionless one-dimensional harmonic oscillator,
\begin{align}
\hat H = -\frac{1}{2}\frac{\dd^2}{\dd x^2} + \frac{1}{2}x^2,
\label{eq:ho_hamiltonian}
\end{align}
whose exact eigenvalues are
\begin{align}
E_n = n + \frac{1}{2}, \qquad n=0,1,2,\ldots .
\label{eq:ho_exact_energy}
\end{align}
Appendix~\ref{app:dimensionless} derives Eq.~(\ref{eq:ho_hamiltonian}) from the dimensional Schr"odinger operator by scaling position with the oscillator length $\ell_0=\sqrt{\hbar/(m\omega)}$ and energy with $\hbar\omega$. Exact eigenfunctions are the Hermite functions,
\begin{align}
\psi_n(x)=\frac{1}{\sqrt{2^n n!\sqrt{\pi}}}H_n(x)\ee^{-x^2/2}.
\label{eq:hermite_function}
\end{align}
These formulas provide an unusually clean calibration target for AI-assisted derivation and implementation because both spectral and wavefunction errors can be measured directly.

\subsection{Finite-difference realization and verification}
The numerical realization uses a centered second-order finite difference on $x\in[-x_{\max},x_{\max}]$ with homogeneous boundary truncation. On interior grid points $x_i$, the discrete Hamiltonian is
\begin{align}
(H\bm\psi)_i = -\frac{1}{2}\frac{\psi_{i+1}-2\psi_i+\psi_{i-1}}{\Delta x^2} + \frac{1}{2}x_i^2\psi_i.
\label{eq:ho_fd}
\end{align}
The resulting tridiagonal matrix is assembled sparsely and diagonalized for the lowest few eigenpairs.

\begin{figure*}[t]
    \centering
    \includegraphics[width=0.98\textwidth]{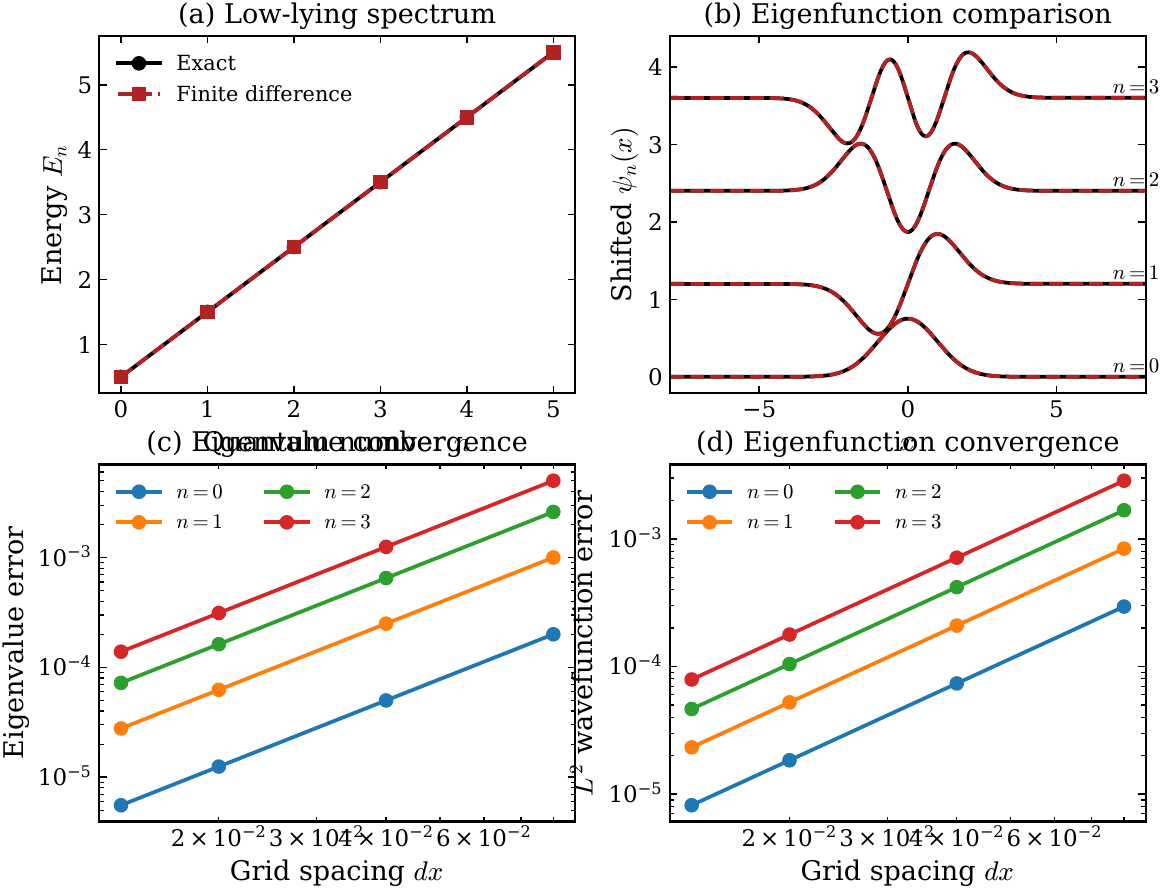}
    \caption{Validation of the harmonic-oscillator case study. (a) Exact eigenvalues $E_n=n+1/2$ (black solid circles) and finite-difference results on the finest grid (red dashed squares). (b) Exact Hermite-function eigenstates (black solid) and numerical eigenvectors (red dashed), shown with vertical offsets for $n=0,1,2,3$. (c) Eigenvalue errors versus grid spacing for the first four levels. (d) Discrete $L^2$ errors of the corresponding eigenfunctions versus grid spacing. The figure demonstrates that the AI-assisted implementation reproduces a canonical quantum benchmark with the correct second-order convergence.}
    \label{fig:oscillator}
\end{figure*}

Figure~\ref{fig:oscillator} reports the validation. Panel (a) compares exact and numerical low-lying eigenvalues on the finest grid. Panel (b) compares representative eigenfunctions after sign alignment and discrete normalization. Panels (c) and (d) show grid-refinement studies for eigenvalue and eigenfunction errors. On the finest grid used in the manuscript ($\Delta x\approx 1.33\times 10^{-2}$), the largest absolute error among the first six eigenvalues is $3.39\times 10^{-4}$, and both eigenvalue and eigenfunction errors follow the expected second-order trend. The fitted log-log slopes for $n=0,1,2,3$ are approximately $2.00$ in both observables.

This example illustrates a narrow but important form of AI usefulness. AI can help derive the dimensionless model, draft the operator assembly, and organize the benchmark comparison. None of those steps is sufficient on its own. The result becomes trustworthy only because the exact spectrum and Hermite functions immediately expose implementation mistakes.

\section{Case study II: parabolic and elliptic PDE validation}
\subsection{Heat equation with an exact modal solution}
The first PDE benchmark is the one-dimensional heat equation
\begin{align}
\partial_t u(x,t)=D\,\partial_x^2u(x,t),\qquad 0<x<L,
\label{eq:heat_pde}
\end{align}
with homogeneous Dirichlet boundary conditions and initial data
\begin{align}
u(x,0)=\sin\!\left(\frac{\pi x}{L}\right)+0.3\sin\!\left(\frac{3\pi x}{L}\right).
\label{eq:heat_ic}
\end{align}
Because the initial state is a short modal expansion, the exact solution is available in closed form,
\begin{align}
u(x,t)=\ee^{-D(\pi/L)^2 t}\sin\!\left(\frac{\pi x}{L}\right)
+0.3\,\ee^{-9D(\pi/L)^2 t}\sin\!\left(\frac{3\pi x}{L}\right).
\label{eq:heat_exact}
\end{align}
The main solver in the manuscript uses Crank--Nicolson time stepping, while an explicit forward-time centered-space (FTCS) discretization is used as a secondary comparison under a stable time-step restriction.

\begin{figure*}[t]
    \centering
    \includegraphics[width=0.98\textwidth]{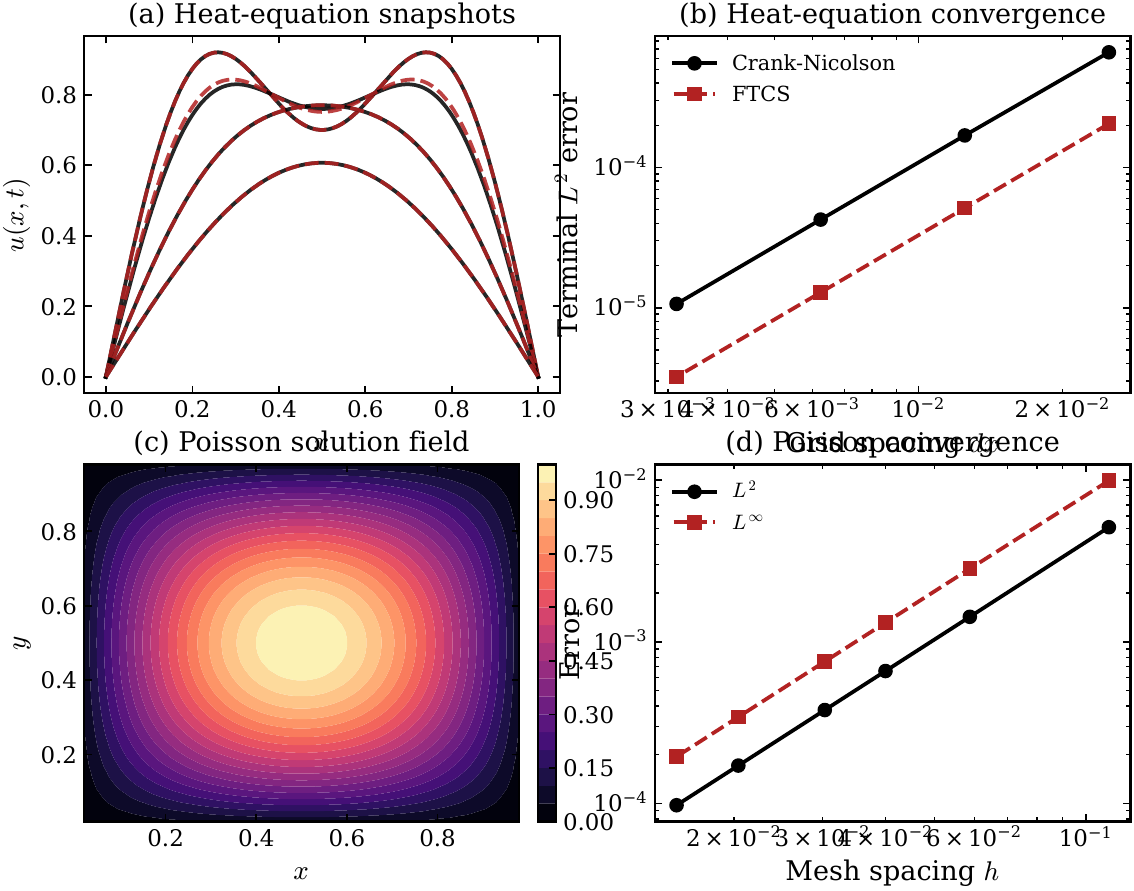}
    \caption{PDE validation results. (a) Heat-equation snapshots at several times, with exact solutions shown by black solid lines and Crank--Nicolson results by red dashed lines. (b) Heat-equation terminal $L^2$ errors versus grid spacing for Crank--Nicolson and FTCS. (c) Numerical solution of the manufactured Poisson problem on a representative grid. (d) Poisson $L^2$ and $L^\infty$ errors versus mesh spacing. The figure demonstrates that the generated solvers agree with exact or manufactured benchmarks and exhibit the expected second-order behavior.}
    \label{fig:pde}
\end{figure*}

The question here is not whether the heat equation can be solved numerically; it obviously can. The relevant question is whether an AI-assisted workflow can assemble a solver, state the stability conditions correctly, produce reproducible convergence data, and document the result without relying on unverifiable claims. Figure~\ref{fig:pde} shows that it can. The Crank--Nicolson and FTCS solutions both converge to the analytic benchmark, and the measured terminal-error slopes are $1.99$ and $2.00$, respectively, under the chosen refinement protocol with $\Delta t\propto \Delta x$ for Crank--Nicolson and $\Delta t\propto \Delta x^2$ for FTCS.

\subsection{Poisson equation with a manufactured solution}
We next solve
\begin{align}
-\nabla^2 u(x,y)=f(x,y)
\label{eq:poisson}
\end{align}
on the unit square with Dirichlet boundary conditions chosen from the manufactured exact solution
\begin{align}
u(x,y)=\sin(\pi x)\sin(\pi y),
\label{eq:poisson_exact}
\end{align}
which implies
\begin{align}
f(x,y)=2\pi^2\sin(\pi x)\sin(\pi y).
\label{eq:poisson_rhs}
\end{align}
The discrete operator is the standard five-point Laplacian, assembled as a sparse Kronecker-sum matrix and solved with a sparse direct method. The refinement study again behaves as expected: the discrete $L^2$ error converges with slope $2.00$, and the pointwise maximum error shows the same second-order trend within finite-grid variation.

This pair of PDE examples is useful because it spans both time-dependent and elliptic problems. AI assistance is strong at boilerplate generation, sparse-matrix assembly, and plotting. Agreement with exact or manufactured benchmarks is what turns that convenience into credible computational science.

\section{Case study III: inverse modeling and uncertainty quantification}
We next consider a standard nonlinear regression problem with synthetic data,
\begin{align}
x(t)=A\ee^{-\gamma t}\cos(\Omega t+\phi)+c+\epsilon(t),
\label{eq:damped_model}
\end{align}
where $\epsilon(t)$ is Gaussian noise with known standard deviation. Synthetic observations are generated with a fixed random seed, ground-truth parameters $(A,\gamma,\Omega,\phi,c)=(1.2,0.35,4.1,0.4,0.1)$, and additive noise level $\sigma=0.08$. Parameters are then recovered by weighted nonlinear least squares, and uncertainty is estimated both from the covariance matrix and from bootstrap refits.

This example is deliberately modest. It is not a benchmark of cutting-edge Bayesian computation. Rather, it asks whether AI assistance can help specify the model, generate synthetic data, fit parameters, diagnose residuals, and attach uncertainty intervals without blurring the distinction between fitted and true values.

\begin{figure*}[t]
    \centering
    \includegraphics[width=0.98\textwidth]{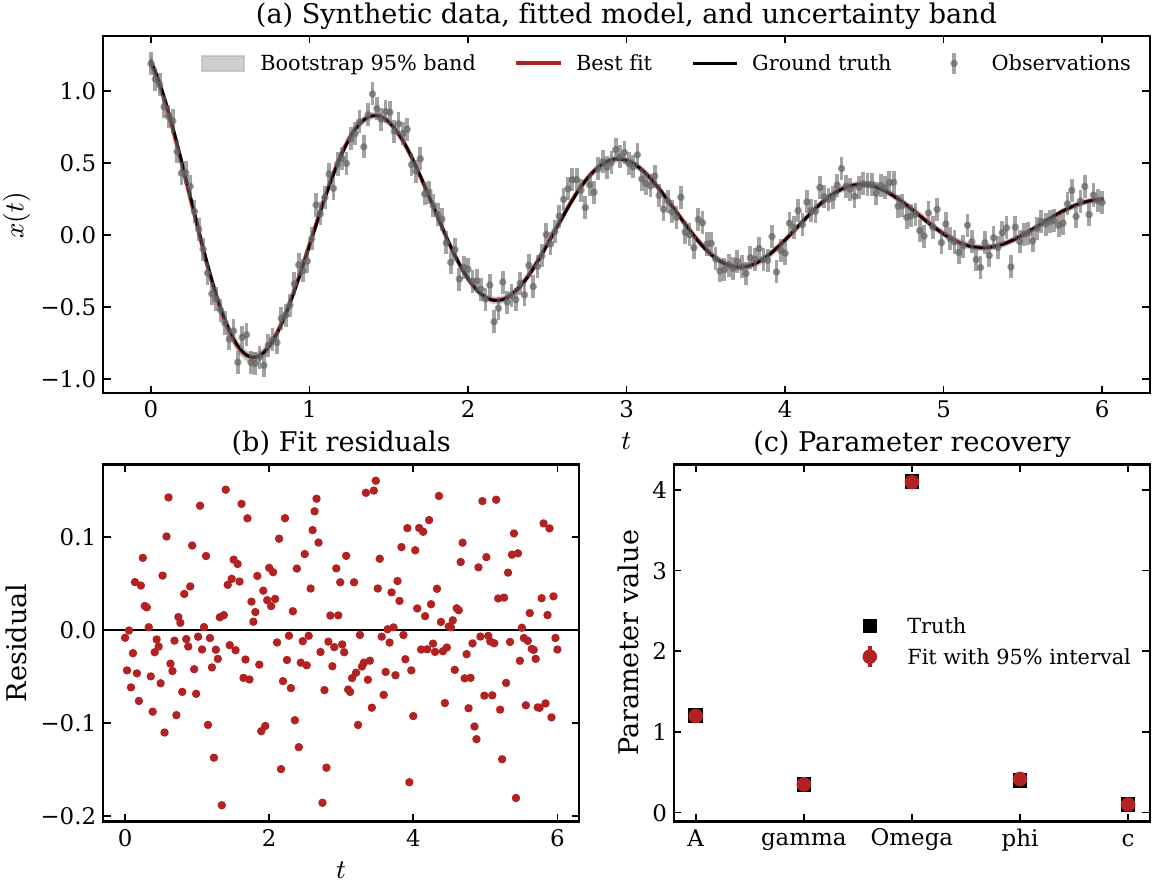}
    \caption{Inverse-modeling demonstration for noisy damped oscillations. (a) Ground-truth signal (black solid), noisy observations with error bars (gray markers), nonlinear least-squares fit (red solid), and bootstrap $95\%$ prediction band (gray shaded region). (b) Residuals of the best fit as a function of time. (c) Fitted parameter values with bootstrap intervals compared with ground truth. The figure shows that the AI-assisted workflow can formulate the inverse problem, run the regression, and present uncertainty information, while still requiring explicit statistical checks.}
    \label{fig:inference}
\end{figure*}

Figure~\ref{fig:inference} shows the outcome. The fitted trajectory follows the data within the imposed noise level, the residuals remain structureless at visual inspection, and the bootstrap band tracks the uncertainty envelope. Table~\ref{tab:fit} reports the recovered parameters. In this fixed-seed realization, the fitted values are close to ground truth; for example, the damping rate is recovered as $\gamma=0.349932$ and the angular frequency as $\Omega=4.094709$. The bootstrap $95\%$ intervals cover all five ground-truth parameters in this realization.

The lesson is again methodological. AI can help build an end-to-end fitting pipeline quickly, but parameter estimates become meaningful only after residual inspection, sensitivity to initialization, and uncertainty quantification are made explicit.

\begin{table}[b]
\caption{Recovered parameters for the damped-oscillation inverse problem. The covariance-based standard deviation is obtained from the least-squares Jacobian approximation, and the bootstrap interval is computed from $250$ residual-resampling refits.}
\label{tab:fit}
\begin{ruledtabular}
\begin{tabular}{lcccc}
Parameter & Truth & Fit & $\sigma_{\mathrm{cov}}$ & Bootstrap 95\% interval \\
\colrule
$A$ & 1.200000 & 1.197717 & 0.024765 & [1.155504, 1.236056] \\
$\gamma$ & 0.350000 & 0.349932 & 0.011111 & [0.331439, 0.368051] \\
$\Omega$ & 4.100000 & 4.094709 & 0.010745 & [4.076847, 4.112104] \\
$\phi$ & 0.400000 & 0.416607 & 0.019095 & [0.384862, 0.453410] \\
$c$ & 0.100000 & 0.100870 & 0.005517 & [0.091422, 0.109997] \\
\end{tabular}
\end{ruledtabular}
\end{table}

\section{Case study IV: algorithmic scaling in scientific computing}
The final case study is comparative rather than analytic. We examine two pairs of algorithms: dense diagonalization versus a sparse eigensolver for the harmonic-oscillator matrix, and sparse direct solve versus conjugate gradients for the Poisson system. The goal is not to produce a universal benchmarking paper. Timing data depend on hardware, libraries, and matrix structure. The point is to demonstrate that AI can help set up a scientifically legible comparison and report its limitations honestly.

\begin{figure*}[t]
    \centering
    \includegraphics[width=0.98\textwidth]{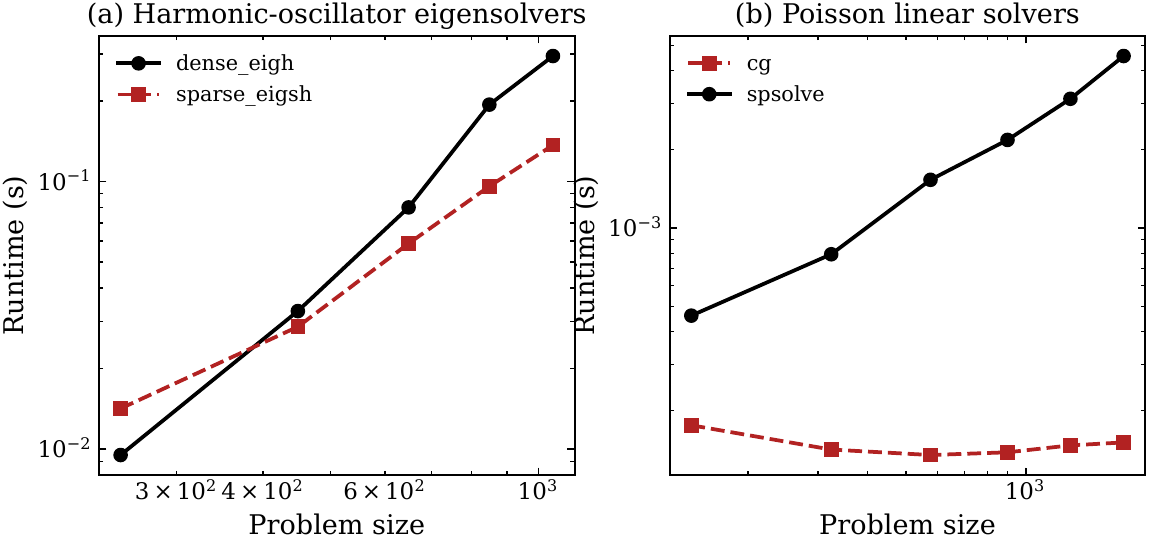}
    \caption{Algorithmic comparisons. (a) Runtime versus problem size for dense diagonalization and sparse low-mode eigensolving of the harmonic-oscillator matrix. (b) Runtime versus problem size for a sparse direct solve and conjugate gradients on the manufactured Poisson problem. Both panels should be interpreted as machine-dependent workflow demonstrations rather than universal performance claims. The main conclusion is that AI can help construct and document algorithmic comparisons in a reproducible form.}
    \label{fig:scaling}
\end{figure*}

For the harmonic oscillator, the dense method computes all eigenpairs of the matrix and is therefore asymptotically more expensive than requesting only a few low modes from a sparse Krylov method. The expected trend appears in Fig.~\ref{fig:scaling}: at smaller sizes the methods are comparable, while at the largest oscillator matrix used here ($N=1049$ interior unknowns) the dense routine is slower by a factor of about $2.15$ for the low-mode task considered. For the Poisson problem, conjugate gradients outperform the sparse direct solve throughout the sampled range while maintaining the same manufactured-solution error to displayed precision because both methods solve the same linear system to tight tolerance.

This section illustrates a useful division of labor. AI assistance is effective for generating benchmarking scaffolding, organizing experiments, and standardizing plots. Human judgment remains necessary for interpreting when a timing comparison is meaningful, when a crossover is expected, and when hardware-dependent results should not be overgeneralized.

\section{Discussion: strengths, limitations, and verification requirements}
The examples above support a restrained but useful conclusion. Present-day AI is already strong at assembling standard derivations, drafting numerical operators, structuring code into modules, producing publication-quality plots, and organizing a manuscript around explicit checks. In that sense, AI can materially reduce friction across large portions of the scientific workflow.

At the same time, the same examples clarify what AI does \emph{not} guarantee. It does not guarantee correct nondimensionalization, correct boundary handling, correct sparse-matrix assembly, correct uncertainty interpretation, or even internally consistent prose. These failures may be silent. For reasoning tasks in particular, current systems can fail to self-correct reliably without external constraints \cite{Huang2024}. Accordingly, trustworthy use requires exact benchmarks, manufactured solutions, convergence studies, sensitivity checks, fixed random seeds where appropriate, and explicit artifact generation.

The present project therefore suggests a practical workflow principle. AI should be treated neither as an oracle nor as a mere conversational convenience. It is best treated as a copilot whose output must continually encounter independent structure: known theory, numerical checks, and reproducibility tooling. This should not be interpreted as implying AI authorship or autonomous scientific judgment. When such structure is present, the productivity gains are real. When it is absent, the risk of polished but incorrect work rises sharply.

There are also limits to the scope of this manuscript. The paper does not test frontier scientific creativity, autonomous hypothesis generation, or laboratory automation. It does not compare AI models against one another. It does not establish that the workflow generalizes unchanged to strongly nonlinear multiphysics or experimental data with serious model mismatch. Those are important questions, but they are different questions. The current paper is a deliberately conservative demonstration on canonical benchmarks.

\section{Conclusion}
We have presented a fully reproducible demonstration paper showing that contemporary AI can already be highly useful across symbolic derivation, numerical simulation, inverse modeling, uncertainty quantification, algorithmic comparison, figure production, and manuscript preparation. The scientific examples were intentionally canonical rather than novel: a harmonic oscillator with exact eigenpairs, a heat equation with a known modal solution, a Poisson problem with a manufactured solution, and a damped-oscillation fitting task with synthetic data and bootstrap uncertainty estimates.

The main result is therefore methodological. AI assistance becomes scientifically valuable when it is paired with benchmark problems, exact checks, convergence studies, quantitative residual analysis, and transparent artifact generation. Under those conditions, AI can function as a practical research copilot. Without those conditions, convenience can easily be mistaken for correctness.

\appendix

\section{Dimensionless conventions and notation}
\label{app:dimensionless}
This appendix states the scaling conventions used throughout the manuscript.

For the dimensional harmonic oscillator,
\begin{align}
\hat H_X=-\frac{\hbar^2}{2m}\frac{\dd^2}{\dd X^2}+\frac{1}{2}m\omega^2X^2,
\end{align}
we define the oscillator length $\ell_0=\sqrt{\hbar/(m\omega)}$ and dimensionless coordinate $x=X/\ell_0$. Because
\begin{align}
\frac{\dd}{\dd X}=\frac{1}{\ell_0}\frac{\dd}{\dd x},
\end{align}
one obtains
\begin{align}
\hat H_X
&=-\frac{\hbar^2}{2m\ell_0^2}\frac{\dd^2}{\dd x^2}+\frac{1}{2}m\omega^2\ell_0^2 x^2 \\
&=\hbar\omega\left(-\frac{1}{2}\frac{\dd^2}{\dd x^2}+\frac{1}{2}x^2\right).
\end{align}
Dividing by $\hbar\omega$ yields Eq.~(\ref{eq:ho_hamiltonian}).

For the heat equation, the manuscript uses $L=1$ and $D=1$ in the generated data, but the formulas are presented for general $L$ and $D$. For the Poisson problem, all coordinates are dimensionless on $[0,1]^2$. The inverse-modeling example treats time and displacement as dimensionless synthetic units, which is sufficient because the point of that case study is statistical verification rather than dimensional analysis.

\section{Harmonic-oscillator derivation and exact benchmark formulas}
\label{app:ho_exact}
This appendix records the exact formulas used to validate the spectral computation.

Introducing the standard ladder operators,
\begin{align}
\hat a = \frac{1}{\sqrt{2}}\left(x+\frac{\dd}{\dd x}\right), \qquad
\hat a^{\dagger}=\frac{1}{\sqrt{2}}\left(x-\frac{\dd}{\dd x}\right),
\end{align}
one finds
\begin{align}
\hat H = \hat a^{\dagger}\hat a + \frac{1}{2}.
\end{align}
Therefore the exact spectrum is $E_n=n+1/2$. The normalized eigenfunctions are given by Eq.~(\ref{eq:hermite_function}), where $H_n(x)$ denotes the physicists' Hermite polynomial. The code evaluates $H_n$ numerically using standard special-function routines and aligns the sign of each discrete eigenvector with the exact benchmark before plotting.

\section{Finite-difference discretization of the Schr\"odinger operator}
\label{app:ho_fd}
This appendix gives the discrete operator used in the harmonic-oscillator solver.

On a uniform grid $x_i=-x_{\max}+i\Delta x$, the second derivative is approximated by
\begin{align}
\frac{\dd^2\psi}{\dd x^2}(x_i)=\frac{\psi_{i+1}-2\psi_i+\psi_{i-1}}{\Delta x^2}+\mathcal{O}(\Delta x^2).
\end{align}
Substituting into Eq.~(\ref{eq:ho_hamiltonian}) gives Eq.~(\ref{eq:ho_fd}), which corresponds to a tridiagonal matrix with entries
\begin{align}
H_{ii}=\frac{1}{\Delta x^2}+\frac{1}{2}x_i^2,
\qquad
H_{i,i\pm1}=-\frac{1}{2\Delta x^2}.
\end{align}
The code stores this operator in sparse format and extracts the lowest modes by a sparse eigensolver. The verification metrics are the absolute eigenvalue error and the grid-weighted discrete $L^2$ error of the eigenfunction,
\begin{align}
\|\psi_h-\psi\|_{\Ltwo,h}=\left(\Delta x\sum_i|\psi_{h,i}-\psi(x_i)|^2\right)^{1/2}.
\end{align}

\section{Heat-equation exact solution and time-stepping details}
\label{app:heat_details}
This appendix provides the details of the parabolic benchmark.

With homogeneous Dirichlet boundary conditions, the sine modes are eigenfunctions of the Laplacian. Because the initial condition is a linear combination of the first and third sine modes, the exact solution is the corresponding superposition of decaying exponentials, Eq.~(\ref{eq:heat_exact}).

For Crank--Nicolson, let $u_i^n\approx u(x_i,t_n)$ and define $r=D\Delta t/\Delta x^2$. The interior update can be written as
\begin{align}
\left(I+\frac{r}{2}L\right)\bm u^{n+1}=\left(I-\frac{r}{2}L\right)\bm u^n,
\end{align}
where $L$ is the standard positive tridiagonal Laplacian matrix with diagonal $2$ and off-diagonal $-1$. This scheme is second-order accurate in both time and space for smooth solutions. The FTCS comparison uses
\begin{align}
u_i^{n+1}=u_i^n+r\left(u_{i-1}^n-2u_i^n+u_{i+1}^n\right),
\end{align}
with a stable time-step choice satisfying $r\le 1/2$.

\section{Poisson manufactured solution and discrete operator construction}
\label{app:poisson_details}
This appendix states the elliptic discretization used in the manuscript.

On an $n\times n$ interior grid with spacing $h=1/(n+1)$, the five-point Laplacian gives
\begin{align}
-\nabla_h^2 u_{i,j}=\frac{4u_{i,j}-u_{i+1,j}-u_{i-1,j}-u_{i,j+1}-u_{i,j-1}}{h^2}.
\end{align}
In matrix form,
\begin{align}
A = I\otimes T + T\otimes I,
\end{align}
where $T$ is the one-dimensional tridiagonal matrix with diagonal $2/h^2$ and off-diagonal $-1/h^2$. The manufactured exact solution is Eq.~(\ref{eq:poisson_exact}), and the corresponding forcing is Eq.~(\ref{eq:poisson_rhs}). Dirichlet data are zero on the boundary because the exact solution vanishes there.

\section{Error norms, convergence definitions, and verification metrics}
\label{app:error_metrics}
This appendix defines the quantitative checks used throughout the paper.

For a grid function $v_h$ and exact sample $v$, the weighted discrete $L^2$ norm in one dimension is
\begin{align}
\|v_h-v\|_{\Ltwo,h}=\left(\Delta x\sum_i|v_{h,i}-v_i|^2\right)^{1/2},
\end{align}
and in two dimensions it becomes
\begin{align}
\|v_h-v\|_{\Ltwo,h}=\left(h^2\sum_{i,j}|v_{h,ij}-v_{ij}|^2\right)^{1/2}.
\end{align}
The discrete maximum norm is
\begin{align}
\|v_h-v\|_{\Linf,h}=\max_{i,j}|v_{h,ij}-v_{ij}|.
\end{align}
Measured convergence rates are obtained from least-squares slopes on log-log data. The verification script checks that the harmonic-oscillator fine-grid errors remain small, that the heat and Poisson convergence slopes exceed $1.8$, that the bootstrap intervals cover the synthetic truths in the fixed-seed inference example, and that the dense-versus-sparse oscillator timing ratio on the largest tested matrix exceeds unity by a clear margin.

\section{Nonlinear least-squares objective and uncertainty-estimation details}
\label{app:inference_details}
This appendix summarizes the inverse-modeling implementation.

The weighted least-squares objective is
\begin{align}
\chi^2(\bm\theta)=\sum_{k=1}^{N_t}\frac{\left[y_k-f(t_k;\bm\theta)\right]^2}{\sigma^2},
\end{align}
with parameter vector $\bm\theta=(A,\gamma,\Omega,\phi,c)$. The initial guess is obtained from simple summary statistics and a frequency estimate from the discrete Fourier transform of the demeaned observations. Bounds are imposed to keep the optimization in a physically reasonable region.

Two uncertainty summaries are reported. The first is the covariance approximation returned by the nonlinear least-squares solver. The second is a residual bootstrap: residuals from the best fit are resampled with replacement, added back to the fitted curve, and the fit is recomputed. The resulting empirical parameter distribution provides the bootstrap confidence intervals shown in Table~\ref{tab:fit} and Fig.~\ref{fig:inference}.

\section{Algorithmic benchmarking protocol and timing methodology}
\label{app:scaling_details}
This appendix records the benchmarking protocol used for the scaling figure.

For the harmonic oscillator, the benchmark compares a dense full eigendecomposition against a sparse Krylov method that requests only the six lowest eigenvalues. For the Poisson problem, the comparison is between a sparse direct factorization and conjugate gradients with a tight relative tolerance. The timed quantity is wall-clock runtime measured by Python's high-resolution timer. For each problem size, the code records the best time among a small number of repeats in order to reduce transient noise.

These timing results should not be interpreted as architecture-independent truths. Their function in the manuscript is narrower: to demonstrate that AI can help set up a reproducible benchmarking scaffold, record both cost and accuracy proxy metrics, and present a cautious interpretation that respects algorithmic context.

\section{Reproducibility manifest and file-generation logic}
\label{app:manifest}
This appendix documents how the artifact stack is generated.

The provenance of the submission is as follows. The initial project stack was generated from one single user prompt that specified the scientific program, validation requirements, deliverables, and writing constraints. The publication-ready version then underwent human review and limited front-matter revision to align authorship and disclosure with standard submission practice.

The top-level script \texttt{run\_all.py} creates all data files, then all figure files, then runs an independent verification script, and finally compiles the manuscript. The data-generating modules are \texttt{oscillator\_demo.py}, \texttt{heat\_demo.py}, \texttt{poisson\_demo.py}, \texttt{inference\_demo.py}, and \texttt{algorithms\_demo.py}. Shared utilities and exact formulas are stored in \texttt{common.py} and \texttt{symbolic\_tools.py}. Figure assembly is handled by \texttt{make\_figures.py}, and verification is handled by \texttt{verify\_results.py}. The generated numerical outputs are stored as CSV files in the \texttt{data/} directory, the PDF figures are stored in \texttt{figures/}, and the verification report is stored in \texttt{output/verification\_report.txt}.

A concise version of the run logic is:
\begin{lstlisting}[language=Python]
ensure_dirs()
run_oscillator()
run_heat()
run_poisson()
run_inference()
run_scaling()
make_all_figures()
verify()
compile_manuscript()
\end{lstlisting}
This structure is intentionally simple. The main demonstration claim is stronger when every figure and every reported data file can be traced to a transparent, rerunnable script.

\end{document}